\documentclass[11pt]{article}
\usepackage[utf8]{inputenc} 
\usepackage[T1]{fontenc}
\usepackage{amsmath, amssymb, bm}  
\usepackage{graphicx}        
\usepackage{siunitx}         
\usepackage{xcolor}          
\usepackage{authblk}         
\usepackage{geometry}
\usepackage{booktabs}
\title{Numerical Studies for EuPRAXIA@SPARC\_LAB Plasma Beam Driven Working Point}

\author[1,*]{S. Romeo}
\author[1]{A. Del Dotto}
\author[1]{M. Ferrario}
\author[1]{A. Giribono}
\author[2]{A.R. Rossi}
\author[3,4]{G.J. Silvi}
\author[1]{C. Vaccarezza}

\affil[1]{Laboratori Nazionali di Frascati, Via Enrico Fermi 54, 00044 Frascati (Rome), Italy}
\affil[3]{Sapienza University, Piazzale Aldo Moro 5, 00185 Rome, Italy}
\affil[2]{INFN-Milan, Via Celoria 16, 20133 Milan, Italy}
\affil[4]{Sezione di Roma - INFN, Piazzale Aldo Moro 5, 00185 Rome, Italy}
\affil[*]{\texttt{stefano.romeo@lnf.infn.it}}

\date{} 

\begin{document}

\maketitle

\begin{abstract}
   The realization of a plasma based user facility on the model of EuPRAXIA@SPARC\_LAB requires to design a working point for the operation that allows to get an high accelerating gradient preserving a low emittance and low energy spread of the accelerated beam. Such beam is supposed to pilot a soft x-ray free electron laser with a wavelength of 2-\SI{4}{\nano\meter}. In this work several simulation scans are presented, varying at the same time the plasma density and driver-witness separation in order to show that, in a realistic working point for EuPRAXIA@SPARC\_LAB, it is possible to find an ideal compromise for a witness with a peak current >1kA that allows to preserve the energy spread of the core (80\% of the charge) below 0.1\%, while maintaining an accelerating gradient inside the plasma module around  of 1 GV/m. The study is completed with a parametric analysis with the aim of establishing the stability requirements of the RF working point and the plasma channel in order to preserve the energy jitter at the same level of the energy spread.
\end{abstract}
\section{Introduction}
The concept of using particle wakefields for acceleration~\cite{hogan2010plasma} has demonstrated potential as a viable option for the development of a novel class of compact particle accelerators. In this scheme, a high charge electron bunch (> 100 pC) generates a plasma wave whose intense longitudinal electric fields are used to accelerate a properly positioned second bunch, called witness. Significant progress has been made in this field through highly encouraging experimental results, demonstrating noteworthy advancements in the attainment of high gradient~\cite{blumenfeld2007energy,litos2014high} and preservation of beam quality~\cite{pompili2021energy,pompili2022free}. The next crucial objective is to synergistically combine these findings towards the realization of a dependable plasma-based accelerator. The EuPRAXIA project~\cite{assmann2020eupraxia} has the goal to realize a plasma-based user facility capable to operate a free electron laser. This device is widely known for its exacting requirements with regard to beam quality. Within this project, the section dedicated to beam-driven research is known as EuPRAXIA@SPARC\_LAB~\cite{ferrario2018eupraxia}, which will be established at the Frascati National Laboratories (LNF). The main requirements for a reliable user facility are the optimization of the working point and the analysis of their stability. In beam driven acceleration the working points performance depends on the correct time delay of the bunches and their durations that need to be tuned accordingly to the plasma channel itself. Consequently, the optimization of the working point can be performed by varying these parameters; on the other hand, any shot to shot variation introduces an instability. Since the space-time structure of the train of bunches encloses an high number of parameters, constructing a robust, optimal working point represent a difficult task. Practical experience in plasma acceleration indicates that acceptable stability can be obtained even for sub-optimally adjusted working points. In previous theoretical works~\cite{romeo2018simulation} emerged that the most troubling parameter in a beam driven accelerator is the phase jitter, due to a combination of driver-witness separation jitter and plasma density jitter. It will be shown in this paper how it is possible for several working points to find an optimal "stability valley" for the energy spread just varying these two parameters. A complete opposite scenario will be described for the energy jitter, where it is not possible to find a stable minimum/maximum since the dependence on both parameters is linear. A simplified scaling law will be furnished in order to give an evaluation of the maximum jitter that can be allowed in an EuPRAXIA@SPARC\_LAB-like user facility.
\begin{table}[!t]
   \centering
   \begin{tabular}{l|c|c}
       \toprule
       \textbf{Parameter} & \textbf{Driver}                      & \textbf{Witness} \\
       \midrule
           $Q$& \SI{200}{\pico\coulomb}&\SI{30}{\pico\coulomb} \\ 
           $\sigma_{x}$& \SI{4.6}{\micro\meter}&\SI{1.2}{\micro\meter} \\ 
           $\sigma_{y}$& \SI{8.1}{\micro\meter}&\SI{1.3}{\micro\meter} \\ 
           $\sigma_{z}$& \SI{53}{\micro\meter}&\SI{6.7}{\micro\meter} \\ 
           $\varepsilon_{x}$& \SI{1.78}{\milli\meter \milli\radian}&\SI{0.68}{\milli\meter \milli\radian} \\ 
           $\varepsilon_{y}$& \SI{2.26}{\milli\meter \milli\radian}&\SI{0.66}{\milli\meter \milli\radian} \\ 
           $\mathcal{E}$& \SI{539}{\mega\electronvolt} &\SI{537}{\mega\electronvolt} \\ 
           $\sigma_{\mathcal{E}}$& 0.06\% &0.05\% \\ 
           $I_{peak}$& \SI{0.6}{\kilo\ampere} & \SI{1.7}{\kilo\ampere} \\ 
           \bottomrule
   \end{tabular}
   \caption{Parameters of the beam for the scans in Fig.(\ref{fig:scan}a) and (\ref{fig:scan}b).}
   \label{Bunch1}
\end{table}
\section{Numerical scans}
The numerical scans are performed in the framework of a start-to-end simulation. The input bunch is obtained by means of previous numerical simulations of the whole RF photoinjector, performed with the numerical codes TStep~\cite{youngtstep} and Elegant~\cite{borland2000elegant}, and based on the latest EuPRAXIA@SPARC\_LAB layout~\cite{giribono2018eupraxia}. The numerical scans are performed over four different working points. Plasma channel is simulated by means of the hybrid fluid-kinetic code Architect~\cite{marocchino2016efficient}. The plasma channel profile is constituted of an injection ramp followed by a flat top. The injection ramp has a squared cosine shape rising from null up to the flat top nominal density in 1 cm. The nominal density value is scanned in order to find best performances in a range between 0.9 - \SI{1.4e16}{\per\centi\meter\cubed}. For these evaluations no exit ramp has been considered and the flat top length is variable since the simulations are interrupted as long as the witness bunch reaches the nominal energy for EuPRAXIA beam driven requirements, namely 1GeV. Since the contribution of the tails in the evaluation of energy spread is overwhelming, in all the data analyses a cut is performed over the tails, taking into account the 80\% of witness particles. Any variation on the beam structure at the injection is performed manually, artificially varying the beam charge, length and separation by means of purely numerical operation performed on the outcoming phase space. This manipulation is indeed a strong approximation of the reality since varying any of this parameters requires a major modification of the beam dynamics in the photoinjector, making it impossible to vary each parameter individually. Nevertheless, the individual contribution of each parameter is exactly the investigation goal of the present work. Finer stability analyses that involve full start-to-end simulations for each different working point layout will be object of future works.\\
The first scan is performed employing an already optimized working point, in order to check the configuration robustness against injection phase and plasma density jitters. Driver and witness parameters are reported in Table \ref{Bunch1}.

\begin{table}[!t]
   \centering
    \begin{tabular}{l|c|c}
       \toprule
       \textbf{Parameter} & \textbf{Driver}                      & \textbf{Witness} \\
       \midrule
           $Q$& \SI{200}{\pico\coulomb}&\SI{30}{\pico\coulomb} \\ 
           $\sigma_{x}$& \SI{5.9}{\micro\meter}&\SI{1.5}{\micro\meter} \\ 
           $\sigma_{y}$& \SI{6.7}{\micro\meter}&\SI{1.5}{\micro\meter} \\ 
           $\sigma_{z}$& \SI{55}{\micro\meter}&\SI{7.3}{\micro\meter} \\ 
           $\varepsilon_{x}$& \SI{1.50}{\milli\meter \milli\radian}&\SI{0.81}{\milli\meter \milli\radian} \\ 
           $\varepsilon_{y}$& \SI{2.46}{\milli\meter \milli\radian}&\SI{0.78}{\milli\meter \milli\radian} \\ 
           $\mathcal{E}$& \SI{540}{\mega\electronvolt} &\SI{538}{\mega\electronvolt} \\ 
           $\sigma_{\mathcal{E}}$& 0.12\% &0.07\% \\ 
           $I_{peak}$& \SI{0.4}{\kilo\ampere} & \SI{1.3}{\kilo\ampere} \\ 
        \midrule
           $Q$& \SI{200}{\pico\coulomb}&\SI{30}{\pico\coulomb} \\ 
           $\sigma_{x}$& \SI{4.5}{\micro\meter}&\SI{1.2}{\micro\meter} \\ 
           $\sigma_{y}$& \SI{6.3}{\micro\meter}&\SI{1.3}{\micro\meter} \\ 
           $\sigma_{z}$& \SI{59.6}{\micro\meter}&\SI{5.5}{\micro\meter} \\ 
           $\varepsilon_{x}$& \SI{2.90}{\milli\meter \milli\radian}&\SI{0.59}{\milli\meter \milli\radian} \\ 
           $\varepsilon_{y}$& \SI{5.30}{\milli\meter \milli\radian}&\SI{0.64}{\milli\meter \milli\radian} \\ 
           $\mathcal{E}$& \SI{540}{\mega\electronvolt} &\SI{537}{\mega\electronvolt} \\ 
           $\sigma_{\mathcal{E}}$& 0.09\% &0.06\% \\ 
           $I_{peak}$& \SI{0.5}{\kilo\ampere} & \SI{1.9}{\kilo\ampere} \\ 
        \midrule
           $Q$& \SI{200}{\pico\coulomb}&\SI{50}{\pico\coulomb} \\ 
           $\sigma_{x}$& \SI{4.0}{\micro\meter}&\SI{1.6}{\micro\meter} \\ 
           $\sigma_{y}$& \SI{8.9}{\micro\meter}&\SI{1.6}{\micro\meter} \\ 
           $\sigma_{z}$& \SI{57.9}{\micro\meter}&\SI{6.0}{\micro\meter} \\ 
           $\varepsilon_{x}$& \SI{2.43}{\milli\meter \milli\radian}&\SI{1.13}{\milli\meter \milli\radian} \\ 
           $\varepsilon_{y}$& \SI{2.30}{\milli\meter \milli\radian}&\SI{1.08}{\milli\meter \milli\radian} \\ 
           $\mathcal{E}$& \SI{536}{\mega\electronvolt} &\SI{532}{\mega\electronvolt} \\ 
           $\sigma_{\mathcal{E}}$& 0.16\% &0.11\% \\ 
           $I_{peak}$& \SI{0.6}{\kilo\ampere} & \SI{2.1}{\kilo\ampere} \\ 
           \bottomrule
   \end{tabular}
      \caption{Parameters of the beams for the injection phase scans in Fig.(\ref{fig:scan}c,\ref{fig:scan}f) (top), Fig.(\ref{fig:scan}d,\ref{fig:scan}g) (center) and Fig.(\ref{fig:scan}e,\ref{fig:scan}h) (bottom).}
   \label{Bunch2}%
\end{table}
\begin{figure*}[!tb]
   \centering
   \includegraphics*[width=1\linewidth]{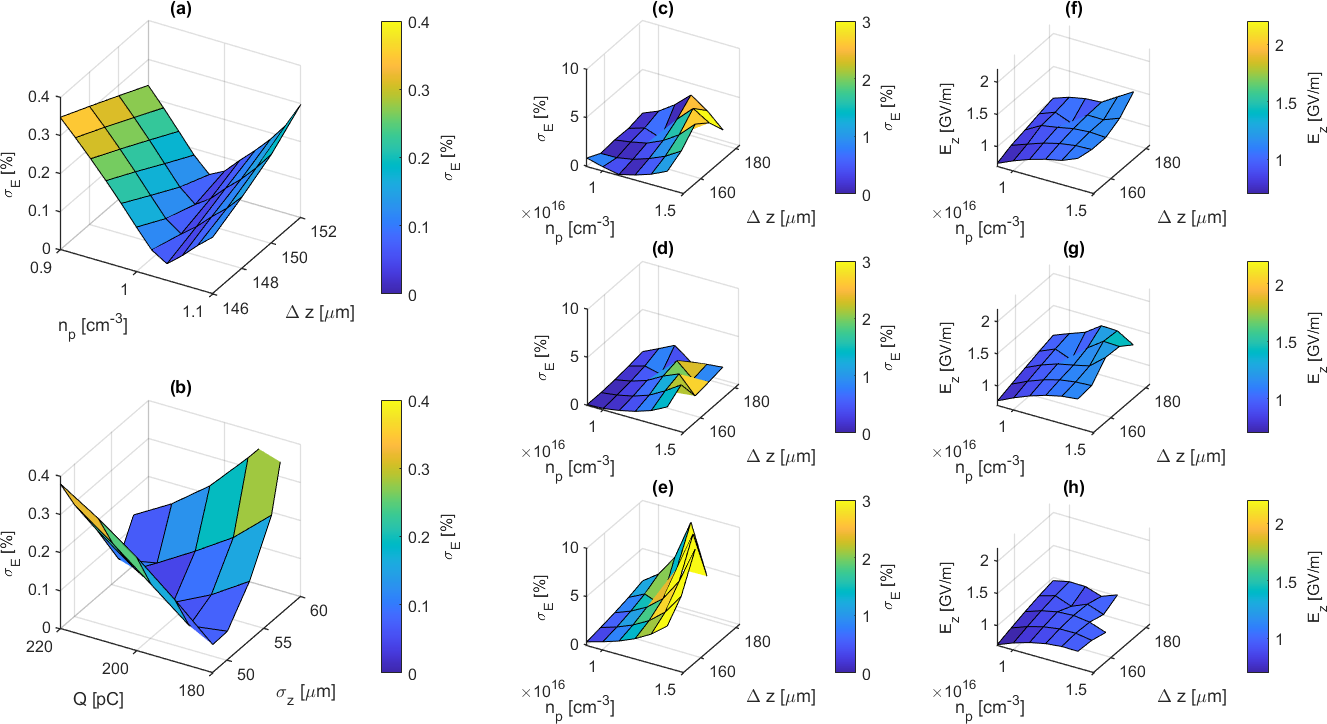}
   \caption{Result of the simulation scans. The first two scans are performed on the working point listed in Table(\ref{Bunch1}) by varying the injection phase (a) and driver length and charge (b). For the working points listed in Table(\ref{Bunch2}) the scan is performed varying the injection phase only. In (c-e) it is shown the witness core energy spread while in (f-h) it is shown the average accelerating gradient.}
   \label{fig:scan}
\end{figure*}
A second scan is performed on the same working point by varying simultaneously the driver length and charge. The final witness energy spread, resulting from both scans, is reported in Fig.(\ref{fig:scan}a,\ref{fig:scan}b). One can note that along a line roughly corresponding to constant phase and current, the energy spread shows a very pronounced minimum with a value around 0.1\%. Despite the growth in energy spread for both cases is of the same order of of magnitude, the charge and length variations employed are way larger than what expected from the RF-photoinjector. On the other hand one can expect a separation jitter of several microns, much higher than the range of \SI{6}{\micro\meter} considered in the analysis and a stability of the density of 10\%, compatible with the average density stability of a plasma module. Based on these observations the following scans are executed by varying the injection phase only, considering three working points with different witness current. The range of density is varied between 0.9 - \SI{1.4e16}{\per\centi\meter\cubed}. Driver and witness parameters are listed in Table \ref{Bunch2}. The results for the energy spread are shown in Fig.(\ref{fig:scan}c-\ref{fig:scan}e). In a similar way to the previous results, a stability region around the minimum energy spread can be identified. This stability valley is realized by the plasma filed flattening due to beam loading effect~\cite{tzoufras2008beam}. Slight changes in the injection phase lead to a non optimal field flattening, resulting in an energy spread increase. Since the energy spread in the stability valley is minimum, all the effects of uneven distribution of energy over the bunch can be neglected because the witness length $\sigma_z$ is much shorter than the plasma wavelength on the plateau $\lambda_p=\sqrt{e^2n_p/c^2\epsilon_0 m_e}$ (where $e$ is the electron charge, $n_p$ the plasma density, $c$ is the speed of light, $\epsilon_0$ is the vacuum permettivity and $m_e$ is the electron mass), i.e. $\sigma_z/\lambda_p \ll 1$. Consequently, one can expect that fluctuations of accelerating field around the stability point for energy spread are linear in phase. In Fig.(\ref{fig:scan}f-\ref{fig:scan}h) this can be verified with a certain degree of approximation. The surface in the left bottom corner of the various plots is approximatively flat, hinting that the average accelerating gradient is linearly proportional to both $n_p$ and $\Delta z$.
\section{Stability requirements}
Assuming the approximation of constant beam loading, one can evaluate a simple expression for the dependency of energy jitter with respect to fluctuations on plasma density and separation jitter. We assume the total blow-out with an immobile ion background in 1 dimensional approximation, where there is no current flowing inside the bubble itself. Gauss law in this approximation becomes
\begin{equation}
    \dfrac{\partial E_z}{\partial z}=\dfrac{e n_p}{\epsilon_0};
    \label{eqn:gauss}
\end{equation}
where $E_z$ is the longitudinal electric field. The electric field $E_z$ results to be a combination of the effects of the blow-out region and the beam loading effect generated by the witness that slightly modifies the shape of the bubble. Nonetheless, Eq.(\ref{eqn:gauss}) is valid before and after the witness, namely that the slope of electric field is constant outside the witness region. We assume perfect beam loading compensation, i.e. consider the field to be flat in the witness region. Said $z_0$ the position of the 0-crossing of the electric field respect to the driver and $\Delta z$ the driver-witness separation, the accelerating gradient is
\begin{equation}
    E_z\approx \dfrac{e n_p}{\epsilon_0} (\Delta z - z_0).
\end{equation}
Now we take small deviations of plasma density and bunch separation respect to the  accelerating gradient acting on the witness and evaluate the relative variation of the electric field
\begin{equation}
    \delta E_z=\dfrac{e}{\epsilon_0}(\Delta z - z_0)\delta n_p-\dfrac{e n_p}{\epsilon_0}\dfrac{\partial z_0}{\partial n_p}+\dfrac{e n_p}{\epsilon_0}\delta z.
\end{equation}
The dependency of the position of the 0-crossing respect to the plasma density will be neglected for sake of simplicity and it will be assumed the approximation that the witness location for realistic working points is close to the bubble trailing area, namely $\Delta z - z_0 \approx \lambda_p/2$. The resulting energy jitter can be written as
\begin{equation}
    \dfrac{\delta \mathcal{E}}{\mathcal{E}-\mathcal{E}_0}=\dfrac{\delta E_z}{E_z}=\dfrac{\delta n_p}{n_p}+\dfrac{2 \delta z}{\lambda_p};
\end{equation}
where $\mathcal{E}$ is the final energy and $\mathcal{E}_0$ is the initial energy. For our setup where $\mathcal{E}=2\mathcal{E}_0$, leading to
\begin{equation}
    \dfrac{\delta \mathcal{E}}{\mathcal{E}}=\dfrac{\delta n_p}{2 n_p}+\dfrac{\delta z}{\lambda_p}.
\end{equation}
Using this equation, and neglecting any possible correlation between $\delta n_p$ and $\delta z$, the stability requirements for an energy jitter of 0.1\% assuming a working point with the characteristics of EuPRAXIA, $\delta n_p\approx$ \SI{2e13}{\per\cubic\centi\meter} and $\delta z \approx$\SI{0.3}{\micro\meter} or equivalently a temporal jitter between driver and witness $\delta t\approx$\SI{1}{\femto\second}. These values are, in principle, the jitters limits from the RF-injector and for the plasma source.
\section{Summary}
In the present work several numerical scans have been performed in order to evaluate the behavior of an optimized particle driven accelerator under different levels of variation of the incoming bunch and plasma profile parameters. A first set of scans varied both driver-witness separation and plasma density. In all the considered cases the existence of a stable local minimum in energy spread emerged, roughly following the path of a constant phase and constant current. Scans of injection phase with different witness current, showed a similar behavior. An equivalent scan of the average electric field acting on witness core has shown a linear dependency over the parameters, hinting the possibility to derive a simplified scaling law for the phase stability of a particle driven accelerator. This scaling law was derived, suggesting that the stability required for a fully operational machine with the EuPRAXIA parameter would require sub-percent stability of plasma density and \unit{\femto\second} scale stability of driver-witness separation . This preliminary result opens a wide scenario for a complete and rigorous stability analysis of the upcoming EuPRAXIA working points.
\section{ACKNOWLEDGEMENTS}
This work was supported by the EuPRAXIA PP grant agreement ID 101079773 and by INFN Commissione Scientifica Nazionale Gruppo V.\\
\bibliographystyle{iopart-num}
\bibliography{biblio}
\end{document}